\documentclass{article}

\usepackage{PRIMEarxiv}

\usepackage[utf8]{inputenc} % allow utf-8 input
\usepackage[T1]{fontenc}    % use 8-bit T1 fonts
\usepackage{hyperref}       % hyperlinks
\usepackage{url}            % simple URL typesetting
\usepackage{booktabs}       % professional-quality tables
\usepackage{amsfonts}       % blackboard math symbols
\usepackage{nicefrac}       % compact symbols for 1/2, etc.
\usepackage{microtype}      % microtypography
\usepackage{lipsum}
\usepackage{fancyhdr}       % header
\usepackage{graphicx}       % graphics
\graphicspath{{media/}}     % organize your images and other figures under media/ folder

\newtheorem{definition}{Definition}

\def \h{\mathbf{h}}
\def \v{\mathbf{v}}
\def \p{\mathbf{p}}

\title{Molecular Property Prediction by Semantic-invariant Contrastive Learning
}

\author{
  Ziqiao Zhang\\
  Fudan University\\
  Shanghai\\
  \texttt{zqzhang18@fudan.edu.cn} \\
   \And
  Ailin Xie \\
  Fudan University\\
  Shanghai\\
  \texttt{alxie21@m.fudan.edu.cn} \\
  \AND
  Jihong Guan \\
  Tongji University \\
  Shanghai \\
  \texttt{jhguan@tongji.edu.cn} \\
  \And
  Shuigeng Zhou$\ast$\\
  Fudan University \\
  Shanghai \\
  \texttt{sgzhou@fudan.edu.cn} \\
}

\begin{document}
\maketitle

\begin{abstract}
    Contrastive learning have been widely used as pretext tasks for self-supervised pre-trained molecular representation learning models in AI-aided drug design and discovery.
    However, exiting methods that generate molecular views by noise-adding operations for contrastive learning may face the semantic inconsistency problem, which leads to false positive pairs and consequently poor prediction performance.
    To address this problem, in this paper we first propose a semantic-invariant view generation method by properly breaking molecular graphs into fragment pairs. Then, 
    we develop a Fragment-based Semantic-Invariant Contrastive Learning (FraSICL) model based on this view generation method for molecular property prediction.
    The FraSICL model consists of two branches to generate representations of views for contrastive learning, meanwhile a multi-view fusion and an auxiliary similarity loss are introduced to make better use of the information contained in different fragment-pair views.
    Extensive experiments on various benchmark datasets show that with the least number of pre-training samples, FraSICL can achieve state-of-the-art performance, compared with major existing counterpart models.

\end{abstract}

% keywords can be removed
\keywords{Molecular representation learning \and Molecular property prediction \and Contrastive learning}

\section{Introduction} 
\label{sec:intro}
Nowadays molecular property prediction (MPP) based on deep learning techniques has been a hot research topic of the AI-aided Drug Discovery (AIDD) community~\cite{gilmer2017neural,duvenaud2015convolutional,jaeger2018mol2vec,song2020communicative,xiong2019pushing,zhang2021fragat,ying2021transformers}.
As most of the molecular properties that drug discovery studies concern require \textit{in vivo} or \textit{in vitro} wet-lab experiments to measure, labeled data for MPP tasks are typically scarce, because it is expensive and time-consuming to acquire such data~\cite{altae2017low}.
On the contrary, there are large amounts of public available unlabeled data~\cite{gaulton2017chembl,10.1093/nar/gkaa971,doi:10.1021/acs.jcim.0c00675}.
Therefore, how to use these large-scale unlabeled molecular data to train deep neural networks to learn better molecular representations for MPP tasks, is of great interest to the AIDD community.

Recently, as self-supervised pre-trained models (e.g. BERT~\cite{devlin2018bert}, MoCo~\cite{he2020momentum} and SimCLR~\cite{chen2020simple}) have shown significant superiority in the fields of Natural Language Processing (NLP) and Computer Vision (CV), self-supervised learning (SSL) has become a mainstream method of utilizing large-scale unlabeled molecular data in MPP study.
These SSL methods typically use some inherent features within or between samples to construct pretext tasks, so that unlabeled data can be leveraged to train deep models in a self-supervised learning manner~\cite{jaiswal2020survey}.
Contrastive learning, masked language model and predictive learning are the currently three categories of methods to design pretext tasks in MPP studies~\cite{wang2022molecular,zhu2021dual,zhu2022featurizations,rong2020self,kim2021merged,xu2021self,hu2019strategies,li2022kpgt}.
Inspired by SimCLR, contrastive learning methods aim at learning representations through contrasting positive data pairs against negative ones~\cite{wang2022molecular}.
Original molecular structures are augmented into multiple views, and views generated from the same molecule are typically used as positive data pairs, while views of different molecules are taken as negative ones~\cite{wang2022molecular}.

% With original molecular structures augmented into multiple views, models are trained by contrastive loss~\cite{bachman2019learning,schroff2015facenet,chen2020simple} to match the representations of positive pairs, i.e. views generated from the same molecule, in the embedding space, and to enlarge the distance between those of negative pairs, i.e.

% Motivate SimCLR~\cite{}, contrastive learning methods first augment original molecular structures to multiple views.
% Views generated from the same molecule are considered as positive pairs, and views of different molecules are negative pairs.
% The goal of contrastive loss~\cite{bachman2019learning,schroff2015facenet,chen2020simple} 
% is to train the model to learn to match the representations of positive pairs in the embedding space, and to enlarge the distance between those of negative pairs~\cite{}.

The way to generate molecular views is crucial to the design of contrastive learning pretext tasks for molecular representation learning.
% To construct contrastive learning tasks, ways to generate views of molecules are essential topics.
As a kind of special objects, molecules can be represented by different methods, including molecular fingerprints~\cite{rogers2010extended}, SMILES~\cite{doi:10.1021/ci00057a005}, IUPAC~\cite{10.1039/9781849733069}, and molecular graph.
These different molecular representation methods therefore can naturally be leveraged to generate views for contrastive learning.
For instance, the DMP~\cite{zhu2021dual} and MEMO~\cite{zhu2022featurizations} models are designed in this way.
Following the practice in CV, another widely used category of methods tries to \textit{add noise} into molecular structures to generate transformations of the original molecules.
These noise-adding operations include deleting atoms, replacing atoms, deleting bonds, deleting subgraph structures etc. 
MolCLR~\cite{wang2022molecular} and GraphLoG~\cite{xu2021self} are such representative models.

\begin{figure}[h]
\begin{center}
\centerline{\includegraphics[width = 6cm]{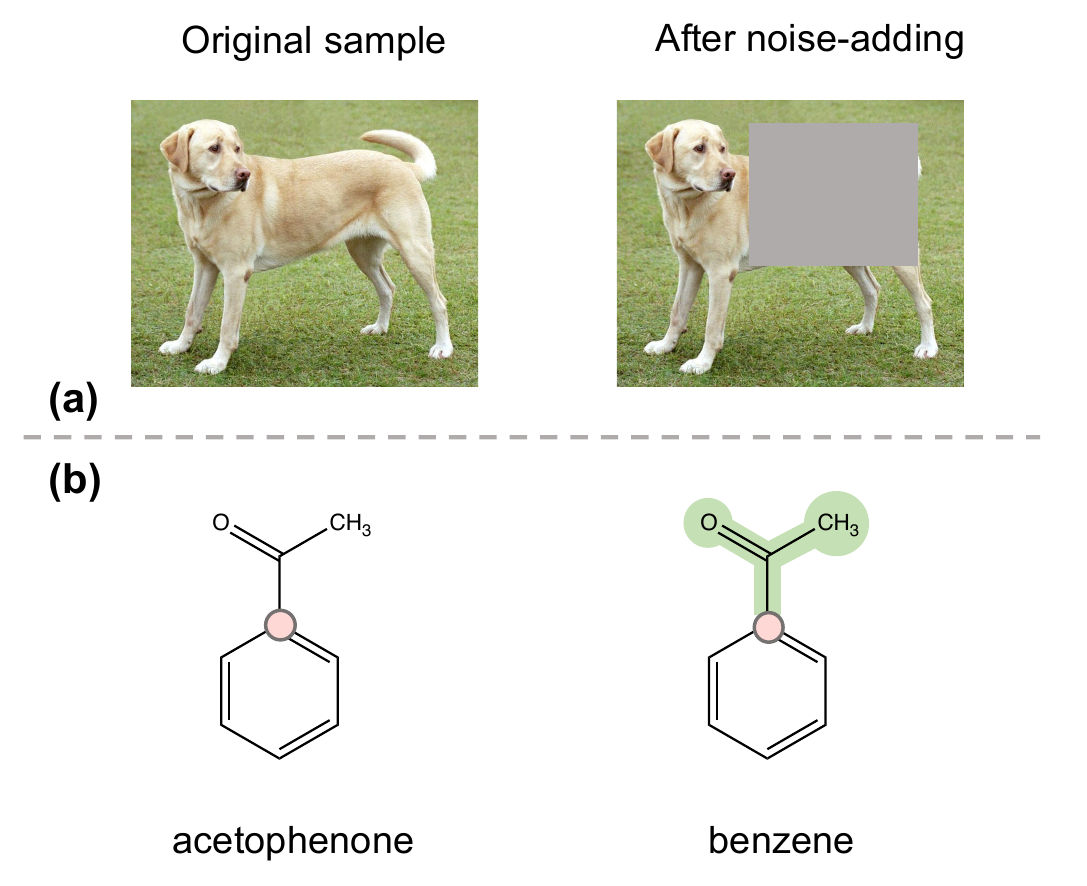}}
\caption{Illustration of the influence of noise-adding operation to the semantic of generated view. (a) After adding noise to the image by randomly masking some area, the semantic of dog image does not change, i.e., the image still represents a dog. (b) By adding noise into the molecular structure by masking some atoms and edges, the molecule \textit{acetophenone}  becomes a completely different molecule \textit{benzene} with different molecular properties. That is, the chemical semantic of \textit{acetophenone} is completely  changed.}
\label{fig:mol_change}
\end{center}
\end{figure}

Although the noise-adding methods for view generation have been widely used in CV studies~\cite{chen2020simple,grill2020bootstrap}, when applying these methods into MPP tasks, a fact that has not been noticed by the researchers is that molecules are very sensitive to noise.
Arbitrarily modifying the topological structure of a molecule with noise, the generated new structure may represent a totally different molecule.
% Arbitrarily changing the topological structure of a molecule, the generated new graph not only can be regarded as a transformation of the original molecule, but also may represent a new molecule.
For instance, as shown in Fig.~\ref{fig:mol_change}(a), adding noise into an dog image by randomly masking some area will not change the semantic of the generated view, which is still a yellow dog.
However, in Fig.~\ref{fig:mol_change}(b), for an acetophenone molecule, deleting a subgraph leads to a benzene molecule, indicating that acetophenone's chemical semantic is completely changed.
And small modification to molecular structure can lead to dramatic changes in the properties of modified molecules, including both bio-activity and other physio-chemical properties.
% However, for a molecule, noise-adding operation (e.g. subgraph deletion in Fig.~\ref{fig:mol_change}) may generate a molecular graph which represents a totally different molecule benzene from the original acetophenone.
% Due to the activity cliff phenomenon in QSAR modeling~\cite{}, small modification to the molecular structure can lead to dramatic changes in the activity of molecules.
% And in addition to bio-acitivity, other physio-chemical properties may also be greatly changed.
Concretely, from the PubChem database we can find that the LogP value of acetophenone in Fig.~\ref{fig:mol_change}(b) is 1.58, while that of benzene is 2.13. The difference is almost 35\%.
Therefore, 
% Therefore, the contrastive view generated by noise-adding operation may represent a completely different molecule with largely different molecular properties, which indicates that a semantic inconsistency phenomenon happens.
it is unreasonable to treat these two views (molecules) as a positive pair for contrastive learning.

Aiming at solving this semantic inconsistency problem, this paper proposes a \textbf{Fra}gment-based \textbf{S}emantic-\textbf{I}nvariant \textbf{C}ontrastive \textbf{L}earning molecular representation model, named FraSICL.
% novel self-supervised pre-trained molecular representation model, named FraContra.
% a contrastive self-supervised pre-trained molecular representation model based on molecular graph fragments are proposed, named FraContra.
A semantic-invariant molecular view generation method is developed, in which a molecular graph is properly broken into fragments by changing the message passing topology while preserving the topological information.
A multi-view fusion mechanism is introduced to FraSICL to make better use of the information contained in views of different fragments and avoid the impact of randomness.
% The FraContra model uses a molecular graph fragmentization method to generate views for contrastive learning.
% In order to avoid the semantic inconsistency problem, the FraContra model proposes a method to break a molecular graph into fragments by changing the message passing topology to generate semantically invariant graph views.
% At the meantime, the model also introduces a multi-view fusion mechanism, which fuses the representations of multiple fragment pairs of a molecule to make a better use of the information contained in the views of different fragments.
% The representation of fused fragment view and the representation of original molecular graph serves as positive pairs for contrastive learning, and the remaining molecules in a batch of inputs are used as negative pairs.
In addition, an auxiliary similarity loss is designed to train the backbone Graph Neural Network (GNN) to generate better representation vectors.

% In addition, the model also uses an auxiliary similarity loss that make the representations of different fragment pairs of the molecule learns from each other to produce better representation vectors.

Our contribution are summarized as follows:
\begin{itemize}
    \item We raise the semantic inconsistency problem in molecular view construction for molecular constrastive learning and develop an effective method to generate semantic-invariant graph views by changing message passing topology while preserving the topological information.
    \item We propose a novel Fragment-based Semantic-Invariant Contrastive Learning molecular representation model for effective molecular property prediction, which is also equipped with a multi-view fusion mechanism and an auxiliary similarity loss to better leverage the information contained in unlabeled pre-training data.
    \item Extensive experiments show that compared with SOTA pre-trained molecular property prediction models, the proposed FraSICL can achieve better prediction accuracy on downstream target tasks with less amounts of unlabeled pre-training data.
\end{itemize}

% \section{Related Works}
% \label{sec:related_works}

% \subsection{Self-supervised learning for molecular property prediction}
% % Self-supervised learning (SSL) have become mainstream methods for utilizing large-scale unlabeled data in MPP study.
% As introduced in Sec.~\ref{sec:intro}, existing SSL methods for MPP studies can be classified into three categories, including contrastive learning, masked language model and predictive learning.
% Contrastive learning aims at learning representations by making the representations between positive pairs closer in the representation space, and enlarge the distances between negative pairs.
% Following the idea of BERT, masked language model methods design pretext tasks by predicting the masked part of the original sample by information of remained parts.
% And predictive learning methods 

% Pre-trained models can be involved into different categories according to their pretext tasks.
% For example, GROVER 

% % \subsection{Molecular graph fragmentation}

\section{Method}\label{sec:method}
Here, we first formally define 
semantic-invariant molecular view in Sec.~\ref{sec:def}, then propose a semantic-invariant molecular view generation method in Sec.~\ref{sec:view_gen} and a multi-view fusion scheme in Sec.~\ref{sec:multi-view}. Finally, 
we present the structure of the Fragment-based Semantic-Invariant Contrastive Learning (FraSICL) molecular representation model in Sec.~\ref{sec:model} and its loss functions in Sec.~\ref{sec:loss}.
% , and the detail of multi-view fusion mechanism and auxiliary similarity loss composing the FraSICL model will be introduced in Sec.~\ref{sec:multi-view} and Sec.~\ref{sec:loss}, respectively.

% In this section, aiming at solving the semantic inconsistency problem in contrastive learning of molecular property prediction models, a novel self-supervised pre-training model FraContra is proposed.
% In Sec.~\ref{sec:def}, the semantic inconsistency problem will be defined formally by introducing two types of semantic inconsistency phenomena.
% Then, a semantic-invariant molecular view generation method is introduced in Sec.~\ref{sec:view_gen}.
% Consequently, the architecture of the proposed FraContra is introduced in Sec.~\ref{sec:model}.
% Multi-view fusion mechanism and auxiliary similarity loss are introduced in Sec.~\ref{sec:multi-view} and Sec.~\ref{sec:loss}, respectively.

\subsection{Semantic-invariant Molecular View}\label{sec:def}
In Sec.~\ref{sec:intro}, we give an example to illustrate how noise-adding operations may lead to semantic inconsistency and consequently false positive pairs.
Here, we will formally define \textit{semantic-invariant molecular view}. 

% to illustrate that adding noise into the structures of molecules to generate new views for contrastive learning may lead to semantic inconsistency problem.
% Here, we will define the semantic inconsistency problem in the molecular representation learning context formally by two types of phenomena, semantic-conflict and semantic-ambiguity.

% In Sec.~\ref{sec:intro}, an example of acetophenone and benzene have been given to illustrate that adding noise into the structure of molecules to generate new views for contrastive learning may lead to semantic inconsistency of the views.
% In this section, the two definitions of semantic inconsistency phenomenon will be proposed formally.

Given a molecule $m$ and its molecular graph $G$ = $\{V,E,X_{atom},X_{bond}\}$ (hydrogen-depleted) where
$V$ denotes the set of nodes that represent the atoms, $E$ denotes the set of edges between nodes, representing the bonds. $X_{atom}$ and $X_{bond}$ are feature matrix of atoms and bonds respectively.
A \textit{transformation function} $F(\cdot)$ is used to generate a {\textit{molecular graph view} (or simply \textit{molecular view}) $G'$ of $G$, i.e.,  $G'$=$F(G)$ and $G'$=$\{V',E',X'_{atom},X'_{bond}\}$. In what follows, we first define two types of semantic inconsistent views.

\begin{definition}[Semantic-conflict view]\label{def:semantic-conflict}
If there is another molecule $m_2$ whose molecular graph is $G_2$, and $G_2$=$G'$=$F(G)$, i.e., the view $G'$ of $m$ is 
%similar to 
the same as
the molecular graph $G_2$ of $m_2$, then we say $G'$ is a \textit{semantic-conflict view} of $m$ with regard to (w.r.t.) $m_2$.
% Assume that the molecular graph of a molecule $m_1$ can be represented by $G_1$, and a function $F(\cdot)$ can transform the molecular graph into a graph view $G'=F(G_1)$.
% If another molecule $m_2$ exists, where $G'$ is the chemical-valid molecular graph of $m_2$, i.e. $G_2 = G'$, then $G'$ is considered to have a semantic-conflict phenomenon to $G_1$.
\end{definition}

\begin{definition}[Semantic-ambiguity view]\label{def:semantic-ambiguity}
If there exists another molecule $m_2$ whose molecular graph is $G_2$, and  $G'_2$=$F(G_2)$=$G'$=$F(G)$, i.e., the view $G'$ of $m$ is
%similar to 
the same as
a view $G'_2$ of $m_2$. Then, we say 
$G'$ is a \textit{semantic-ambiguity view} of molecule $m$ w.r.t. molecule $m_2$.

% Assume that the molecular graph of a molecule $m_1$ can be represented by $G_1 = \{V,E,X_{atom},X_{bond}\}$, and a function $F(\cdot)$ can transform the molecular graph into a graph view $G'=F(G_1)$, $G'=\{V',E',X'_{atom},X'_{bond}\}$.
% If another molecule $m_2$ exists, where its molecular graph $G_2$ can be transformed into $G'$ by the same function $F(\cdot)$, i.e. $G'=F(G_2)$, then $G'$ is considered to have a semantic-ambiguity phenomenon.
\end{definition}

Both semantic-conflict views and semantic-ambiguity views will lead to false positive pairs for molecular representation contrastive learning.
For example, assume that a Graph Neural Network $g(\cdot)$ serves as an encoder to embed the molecular graphs into latent graph embeddings $\h_G=g(G)$.
If we ignore the randomness in the encoder, it is obvious that, for molecule $m$, if it has a semantic-conflict view w.r.t. molecule $m_2$, i.e., $G'$=$F(G)$=$G_2$, then the representation of $G'$ embedded by the graph neural network will be the same as that of $G_2$.
That is, $\h_{G'}$=$g(G')$=$\h_{G_2}$=$g(G_2)$.
% It is obvious that for a given molecular graph $G_1$, if the generated view $G'=F(G_1)$ have a semantic-conflict phenomenon, the representation of $G'$ embedded by the graph neural network $\h'$ will be the same as that of $G_2$, i.e. $\h'=g(G')=g(G_2)=\h_2$, regarding the randomness in the encoder.
In this case, as $\h_{G}$ and $\h_{G'}$ are considered as a positive pair in contrastive learning, $\h_G$ and $\h_{G_2}$ are consequently used as a positive pair.
In another word, the contrastive loss will implicitly make the representations of molecule $m$ and $m_2$ to be close.
However, as claimed before, the molecular properties of different molecules may be greatly different, so that they cannot be used as a positive pair for contrastive learning.
Therefore, semantic-conflict views will lead to false positive pairs and degrade learning performance.

On the other hand, if a semantic-ambiguity view is generated as defined in Def.~\ref{def:semantic-ambiguity}, i.e., $F(G_2)$=$G'$=$F(G)$, 
%as $G_1$ and $G'$ are used as positive pairs in contrastive learning, 
indicating that the contrastive loss will make $\h_G$ and $\h_{G'}$, $\h_{G_2}$ and $\h_{G'}$ to be close in the embedding space, thus $\h_G$ and $\h_{G_2}$ to be close too.
That is, the representations of molecules $m$ and $m_2$ are consequently close by contrastive learning.
% $\h_1$ and $\h'$ are used as positive pairs in contrastive learning, it indicates that the contrastive loss will make the representation of molecule $m_1$ and $m_2$ to be closer.
% it indicates that in the training by contrastive learning, $\h_1$ and $\h'$ are considered as a pair of positive pairs, and $\h_2$ and $\h'$ are also treated as positive pairs.
% Then, $\h_1$ and $\h_2$ are consequently considered as a pair of positive pairs and the embeddings vectors are trained to be closer in the representation space.
So semantic-ambiguity views will also lead to false positive pairs.

To boost the performance of contrastive learning for MPP, we should avoid the generation of both semantic-conflict views and semantic-ambiguity views. That is, we generate only semantic-invariant views, which are defined as follows:

\begin{definition}[Semantic-invariant view]\label{def:semantic-invariant}
Given a view $G'$ of molecule $m$ with graph $G$, if $G'$ is neither a semantic-conflict view nor a semantic-ambiguity view w.r.t. any other molecules, then we say 
$G'$ is a \textit{semantic-invariant view} of $m$.
\end{definition} 

In next section, we will give a method to generate semantic-invariant views.

\subsection{Semantic-invariant View Generation}\label{sec:view_gen}

% In the previous subsection, we have formally defined two types of semantic inconsistency problem, and pointed out that these problems will lead to false positive pairs for contrastive learning.
% Two examples are proposed to illustrate that some noise-adding operation will cause semantic inconsistency problem.
% In this section, a semantic-invariant view generation method will be proposed, which can generate views that can be encoded into different latent representations, but will not lead to semantic-conflict or semantic-ambiguity phenomena.

According to Def.~\ref{def:semantic-invariant} in Sec.~\ref{sec:def}, semantic-invariant views should be neither  
semantic-conflict views nor semantic-ambiguity views. Besides, from the perspective of prediction, they should also be discriminative. That is, they can be encoded into different representations by neural network encoders. 
In this section, to achieve these goals, we propose a semantic-invariant view generation method.

In our previous study~\cite{zhang2021fragat}, to better capture the hierarchical structural information of molecules, a chemical-interpretable molecule fragmentization method FraGAT is proposed.
By considering acyclic single bonds as boundaries between functional groups, the FraGAT model proposes to randomly breaking one of the acyclic single bonds to generate two graph fragments corresponding to some chemical meaningful functional groups.
The experimental results show that learning representations by chemical meaningful molecular graph fragments can achieve good predictive performance for MPP tasks.
% Their experimental results show that learning representations by molecular graph fragments can achieve good predictive performance for MPP tasks.
Inspired by these findings, our semantic-invariant view generation method is designed as follows: 
 
% In the work of FraGAT~\cite{FraGAT}, a molecule fragmentization method is proposed.
% For a given molecule, by randomly breaking one of the acyclic single bonds in the molecular graph, two graph fragments corresponding to some chemical meaningful functional groups are generated.
% Using this disconnected graph with two components as input, the graph neural network will produce a different latent representation with that of the original molecular graph.
% So, these disconnected graphs can serve as views of molecular graph for contrastive learning.

Given a molecule $m$, its molecular graph can be denoted as an annotated graph $G = \{V, E, X_{atom}, X_{bond}\}$.
The atom feature matrix $X_{atom}$ and the bond feature matrix $X_{bond}$ are computed according to Tab.~\ref{tab:feat}.
Then, remove one of the acyclic single bonds $e_{ij}$ from $E$, we obtain $G' = \{V, E', X_{atom}, X_{bond}\}$ where $E'$ = $E - \{e_{ij}\}$.
% Then, when randomly breaking a acyclic single bond in $G$, only the set of edges $E$ is modified.
% The selected edge $e_{ij}$ is deleted from $E$ to obtain $G' = \{V, E', X_{atom}, X_{bond}\}$, where $E = E' \cup \{e_{ij}\}$.
We accept $G'$ as a view to be generated, i.e., a semantic-invariant view.
As the graph $G'$ consists of two disconnected molecular graph fragments, it is also called \textit{fragment-pair view}.

From the discrimination perspective, $G'$ is a different graph from the original molecular graph $G$, so that it will make GNN encoders to generate a different representation. Furthermore, as all the acyclic single bonds in a molecule are unique, breaking different acyclic single bonds will lead to different fragment-pair views, whose representations after a GNN encoder will also be different. That is, the generated views for a molecule are discriminative. 

% On the other hand, to avoid semantic inconsistency problem after breaking the acyclic single bond to form a fragment pair, here we use a method of changing the message passing path of the graph neural network to construct such views.
% Specifically, for a given molecule $m$, the molecular graph can be denoted as $G = \{V, E, X_{atom}, X_{bond}\}$.
% The atomic feature matrix and the bond feature matrix are computed according by the same setting of the FraGAT, as shown in Tab.~\ref{tab:feat}.
% Then, when randomly breaking a acyclic single bond in $G$, only the set of edges $E$ is modified.
% The selected edge $e_{ij}$ is deleted from $E$ to obtain $G' = \{V, E', X_{atom}, X_{bond}\}$, where $E = E' \cup \{e_{ij}\}$.
% At this time, the obtained $G'$ is the view that meets the requirements.

\begin{table}
\centering
\caption{Properties of atoms and bonds in $X_{atom}$ and $X_{bond}$.}
\resizebox{\linewidth}{!}{
\begin{tabular}{cl}
\toprule
Indices of atomic features & Description\\
\midrule
0-15 & Atomic symbol, a one-hot vector of [B,C,N,O,F,Si,P,S,Cl,As,Se,Br,Te,I,At,metal]\\
16-21 & Number of bonds\\
22 & Electrical charge\\
23 & Number of radical electrons\\
24-29 & Hybridization, a one-hot vector of [sp, sp$^2$, sp$^3$, sp$^3$d, sp$^3$d$^2$, other]\\
30 & Aromaticity\\
31-35 & Number of connected hydrogens\\
36 & Whether the atom is a chiral center\\
37-38 & Chirality type, a one-hot vector of [R,S]\\
\midrule
Indices of bond features & Description\\
\midrule
0-3 & Bond type, a one-hot vector of [single, double, triple, aromatic]\\
4 & Whether the bond is conjugated\\
5 & Whether the bond is in a ring\\
6-9 & Stereo, a one-hot vector of [StereoNone, StereoAny, StereoZ, StereoE]\\
\bottomrule
\end{tabular}
}
\label{tab:feat}
\end{table}

% Whether there are semantic inconsistency problem in this method, it will be discussed below.

Then, is $G'$ a real semantic-invariant view of molecule $m$ according to Def.~\ref{def:semantic-invariant}? Let us check.  

On the one hand, as the atom feature matrix $X_{atom}$ is not modified, the numbers of bonds of the two vertex $i$ and $j$ of the broken bond $e_{ij}$ encoded in the atom feature vectors of the generated $G'$ remain the same as that in the original molecular graph $G$.
However, the modified $E'$ indicates that there is no edge between $i$ and $j$, so that the numbers of bonds encoded in the atom feature vectors are not consistent with that in the topological structure.
The degrees of node $i$ and $j$ in graph $G'$ are lower than the numbers of bonds of $i$ and $j$ encoded in $X_{atom}$.
In another word, $G'$ is not a valid molecular graph of any molecule. Furthermore, from the graph perspective, because $G'$ consists of two disconnected subgraphs, and no any valid molecule corresponds to a disconnected graph. 
Therefore, $G'$ cannot be a semantic-conflict view w.r.t. any molecule according to Def.~\ref{def:semantic-conflict}.

% First, in the view $G'$ generated by this method, the atom feature matrix are not modified.
% First, in the view $G'$ generated by this method, neither the number of atoms nor the atom feature matrix are modified.
% As the number of hidden hydrogen atoms and the number of bonds of each atom are encoded in the atom feature matrix, the topological structure of $G'$ represented by the node set $V$ and the edge set $E'$ is inconsistent with the
% According to Tab.~\ref{tab:feat}, we can see that the number of hidden hydrogen atoms and the number of bonds are encoded in the atom feature matrix.
% That is to say, the number of hidden hydrogen atoms and the number of chemical bonds recorded in the feature matrix of the two vertex $i$ and $j$ of the broken bond $e_{ij}$ are inconsistent with the topological structure of $G'$ represented by the node set $V$ and the edge set $E'$.
% In another word, $G'$ does not conform to the chemical rule.
% Therefore, $G' = \{V, E', X_{atom}, X_{bond}\}$ is the molecular graph of any molecules, so that the semantic-conflict problem is avoided.
%
On the other hand, since only one single bond is removed in view $G'$, this discrepancy can only be discovered at nodes $i$ and $j$, and the numbers of bonds of $i$ and $j$ encoded in $X_{atom}$ must be only 1 larger than the degrees of $i$ and $j$, so the removed edge can only be between $i$ and $j$, and the removed edge can only be a single bond.
Thus, there is no other molecular graph $G_2$ that can generate the same $G'$. Similarly, from graph perspective, the graph $G$ of molecule $m$ cannot be equal to the disconnected graph of any view generated from any other molecule. In summary, $G'$ cannot be a semantic-ambiguity view w.r.t. any molecule according to Def.~\ref{def:semantic-ambiguity}.
% And as the 

% On the other hand, since only one edge is removed in view $G'$, in the atomic feature matrix, only the number of chemical bond of two atoms $i$ and $j$ are larger than the actual number of heavy atom neighbors in the topology.
% So, the broken edge can only be between $i$ and $j$, i.e. $e_{ij}$.
% At the same time, since the number of chemical bonds of $i$ and $j$ recorded in the atomic feature matrix is only 1 larger than the actual number of heavy atoms in the topology, the edge between $i$ and $j$ can only be a single bond.
% Therefore, there is no other molecular graph $G_2$ that can generate the same $G'$ by breaking edges, which means that there is no semantic-ambiguity problem.

% For more illustration, the idea of this method can be illustrated from the graph rewiring perspective.
% Specifically, since the information of the nodes of the input graph is propagated according to the edge topology of the input graph in GNNs, changing the edge list $E$ will make the information of nodes be propagated and read out along a different new path.
% And recent studies have pointed out that changing the message passing path of GNNs and decoupling the input graph topology from the message passing topology can effectively alleviate some limiations of GNNs, such as over-smoothing~\cite{}, over-squashing~\cite{}, and heterophily graph learning problem~\cite{}.
% Commonly used dummy node techniques and Graphormer model can also be regarded as methods and models designed based on this idea to improve the learning capability of GNNs.

Finally, from the perspective of graph rewiring~\cite{topping2021understanding,alon2020bottleneck,bruel2022rewiring}, the topological information about the broken edge is encoded in the atom feature matrix $X_{atom}$.
So our method preserves the topological structural information of the original molecular graph, but propagates message between nodes through a different topology.
Thus, it realizes local decoupling of the input graph topology and the message passing topology.
Moreover, compared with randomly breaking any edges in the molecular graph, our method can generate chemical meaningful graph fragments to benefit the prediction of molecular properties, which has been demonstrated in the experiments of previous work~\cite{zhang2021fragat}.

In conclusion, our method is expected to generate better positive pairs, which will help to train neural networks to generate better molecular representations by contrastive learning.

\subsection{Multi-view Fusion}\label{sec:multi-view}

The number of acyclic single bonds of an organic molecule is often large, so there are various fragment-pair views can be generated from one molecule by our proposed view generation method.
As demonstrated in the experiments of some existing work~\cite{zhang2021fragat}, different fragment pairs contain different information about functional groups that constitute a molecule, which shows different predictive performance.
So, to ensure that the information of the functional groups that determine molecular properties and are contained in the fragment pairs can be obtained by the neural network, in FraSICL, we no longer randomly generate \textit{fragment-pair views} as other contrastive learning models do.
Instead, a multi-view fusion mechanism is introduced as follows: 
Given a molecule with $N_b$ breakable acyclic single bonds, all of the $N_b$ fragment-pair views are generated and the representations of these fragment-pair views are calculated by a GNN encoder.
Then, a Transformer encoder is exploited to fuse these representations by the multi-head attention (MHA) mechanism to produce a representation vector that contains information of all of the fragment pairs, named \textit{fragment view}. The details of fragment-pair view fusion are delayed to the next section.  
The fragment view and the molecule view ( i.e., the original molecular graph) are used as two views of a molecule for contrastive learning.

\subsection{Model Structure}
\label{sec:model}

\begin{figure*}[h]
\begin{center}
\centerline{\includegraphics[width = 0.8\linewidth]{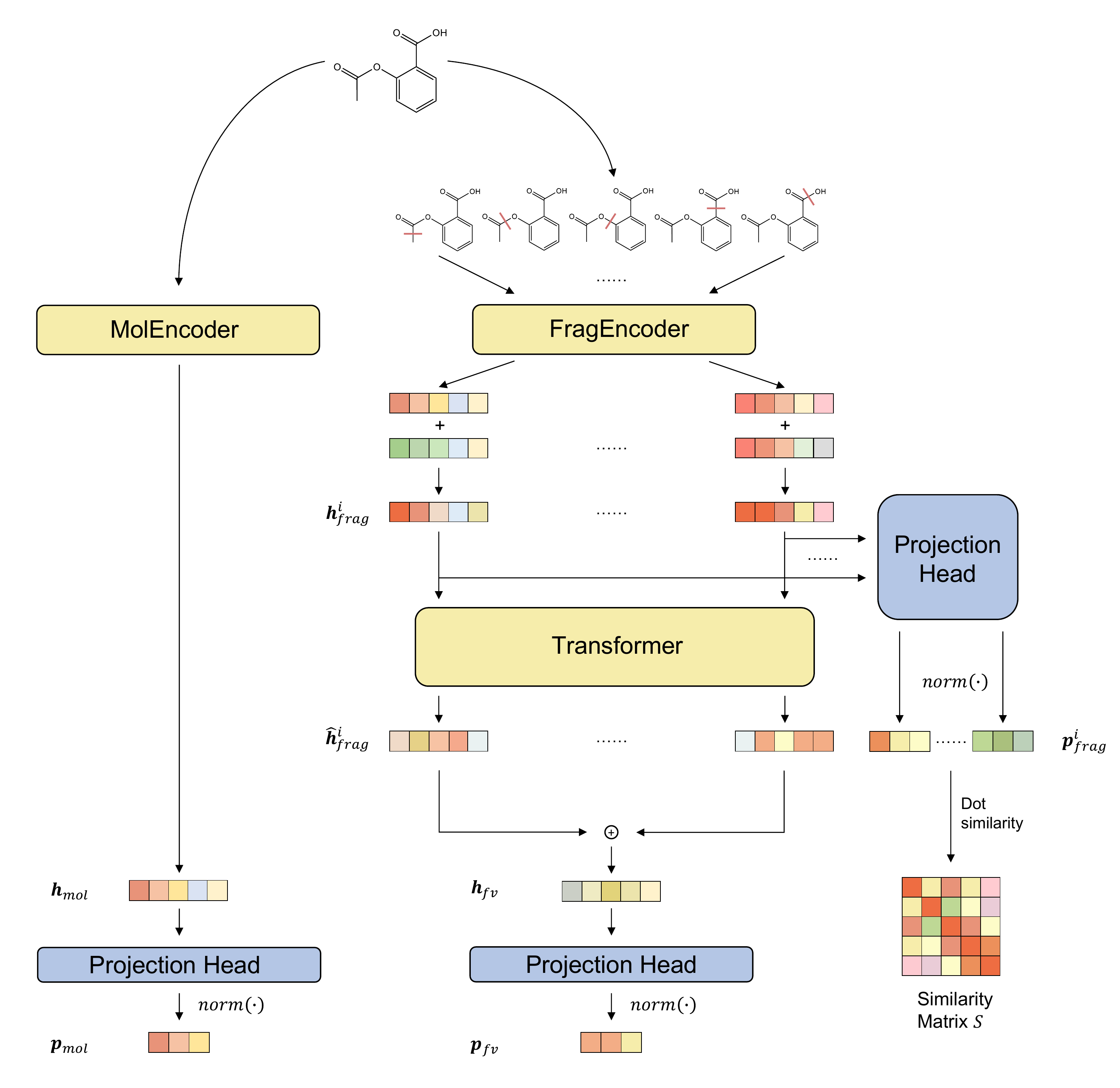}}
\caption{The structure of the FraSICL model.}
\label{fig:FraContra}
\end{center}
\end{figure*}

\begin{figure}[h]
\begin{center}
\centerline{\includegraphics[width = 4cm]{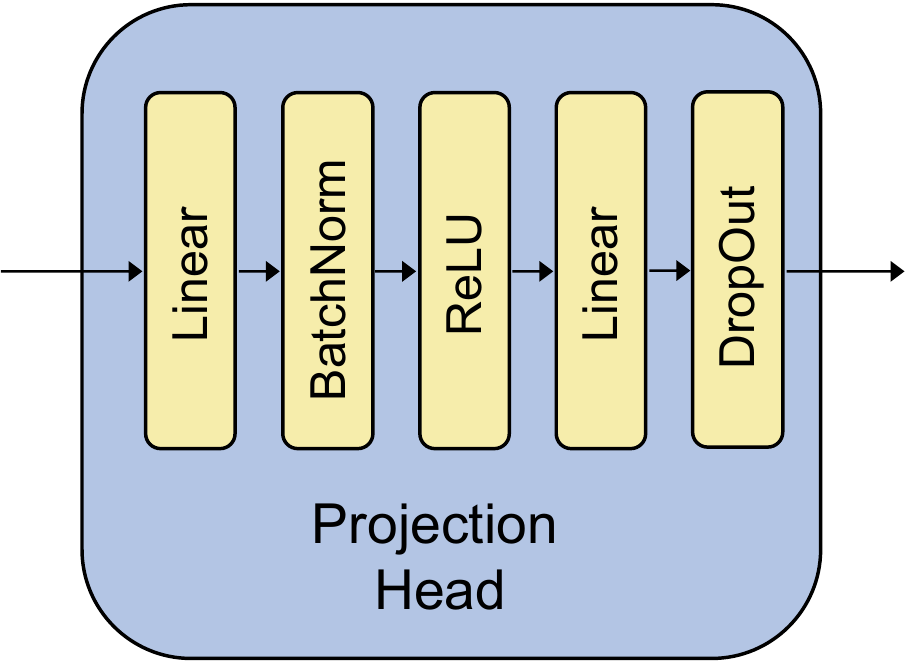}}
\caption{The structure of a projection head. Following the design proposed in BYOL~\cite{grill2020bootstrap}, a projection head consists of a stack of linear layer, BN layer, activation layer, linear layer and dropout layer.}
\label{fig:ProjectHead}
\end{center}
\end{figure}

The structure of the FraSICL model is shown in Fig.~\ref{fig:FraContra}.
Given a molecule $m$ with molecular graph $G_{mol}=\{V, E, X_{atom}, X_{bond}\}$, 
% where $V$ represents the set of atoms in the molecular graph (excluding hydrogen atoms), $E$ represents the set of chemical bonds between atoms in the molecular graph, $X_{atom}$ is the atomic feature matrix and $X_{bond}$ is the bond feature matrix.
% Features of atoms and bonds are computed according to Tab.~\ref{tab:feat}.
the model computes the representations of two views via two branches: the left branch is the \textit{molecule view branch} for generating molecular view, and the right one is the \textit{fragment view branch} for generating the fragment view.

In the molecule view branch, a GNN $g_{mol}(\cdot)$ is used as an encoder to capture the representation of the molecular graph $\h_{mol}=g_{mol}(G_{mol})$.
Attentive FP~\cite{xiong2019pushing} is employed as the graph encoder in this work.
% In this work, the Attentive FP model~\cite{} is used as a graph neural network encoder.
Then, following the structure of MolCLR~\cite{wang2022molecular}, $\h_{mol}$ is fed to a projection head $l_{mol}(\cdot)$ and a regularization function $norm(\cdot)$ to produce the projection of the molecule view $\p_{mol}=norm(l_{mol}(\h_{mol}))$.
The structure of a projection head is shown in Fig.~\ref{fig:ProjectHead}.
And the regularization function is $norm(\v) = \frac{\v}{\|\v\|}$, which can make the length of the projection vector be 1.
% , which is convenient for calculating the projection similarity in contrastive loss.

And for the fragment view branch on the right, all of the $N_b$ breakable acyclic single bonds are enumerated and broken by the method proposed in Sec.~\ref{sec:view_gen} to generate $N_b$ fragment-pair views $G_{frag}^1$=$\{V,E_1,X_{atom},X_{bond}\}$, $\dots$, $G_{frag}^{N_b}$=$\{V,E_{N_b},X_{atom},X_{bond}\}$.
Then, a GNN $g_{frag}(\cdot)$ is used as an encoder to compute the representation of each fragment-pair view $\h_{frag}^i=g_{frag}(G_{frag}^{i})$.
% And for the fragment view branch on the right, first, list out the $N_b$ breakable edges of the given molecule, and break these breakable bonds according to the method proposed in Sec.~\ref{sec:view_gen} to generate $N_b$ fragment-pair view $G_{frag}^1=\{V,E_1,X_{atom},X_{bond}\}$, $\dots$, $G_{frag}^{N_b}=\{V,E_{N_b},X_{atom},X_{bond}\}$.
% Then, using a graph neural network $g_{frag}(\cdot)$ as an encoder, the representation vector of each fragment pair view $G_{frag}^i$ is calculated, $\h_{frag}^i = g_{frag}(G_{frag}^{i})$.
Attentive FP is also used here.
Note that since there are two disconnected components in each fragment-pair view $G_{frag}^i$, $g_{frag}(\cdot)$ will read out these two subgraphs separately and produce two subgraph embeddings.
The representation of a fragment-pair view is obtained by element-wisely adding its two corresponding subgraph embeddings for permutation-invariant property.
% This is different from the approach in FraGAT where a concatenation is used.
% The summation used here can make the model permutation-invariant, so that the output representation will not change when the order of these two graph embeddings changes.

Then, as described in Sec.~\ref{sec:multi-view}, a multi-view fusion mechanism is introduced for leveraging all of the information related to functional groups contained in the $N_b$ fragment-pair views.
% Afterwards, as described in Sec.~\ref{sec:multi-view}, as all of the $N_b$ fragment-pair views contain important information related to the functional groups, and in order to avoid the influence of randomness and the computation complexity caused by the excessive number of positive pairs, a Transformer encoder module is used in the FraContra model to weighted fuse the embedding of $\h_{frag}^i$ of different fragment-pair views.
Specifically, a Transformer encoder $T(\cdot)$ is employed, which uses the representations of fragment-pair views $\h_{frag}^i$ as input tokens, and computes the interaction relationships between the fragment-pair views by the multi-head attention (MHA) mechanism.
The resulting attention scores serve as weights to fuse the representations and obtain $\hat{\h}_{frag}^i$.
% Following the Graphormer~\cite{}, a variation of Transformer encoder is leveraged here, where the last regularization layer (Layer Normalization) of each encoder layer is placed at the beginning of this layer, since this variation is proved to have a more efficient optimization performance~\cite{narang2021transformer}.
By summing up all of the representations of $N_b$ \textit{fragment-pair views}, we can get the representation of \textit{fragment view} $\h_{fv} = \sum_{i=1}^{N_b}\hat{\h}_{frag}^i$.
% The representation of fragment view contains the information of all $N_b$ fragment pairs and the information embedded in the representations of fragment-pair views are fused through MHA mechanism, therefore the generated representation is expected to 
% The representation of fragment view contains the information of all $N_b$ fragment pairs that can be obtained from a molecule, and the information embedded in the representation of the fragment-pair views are fused through MHA mechanism.
% So that the model can focus on important fragment pairs, which can hopefully generate better representation vectors.
Finally, the representation $\h_{fv}$ of the fragment view goes through a projection head $l_{fv}(\cdot)$ and a normalization layer to get $\p_{fv} = norm(l_{fv}(\h_{fv}))$.
Following the structure of MolCLR, the two projections $\p_{mol}$ and $\p_{fv}$ are used to calculate contrastive loss.
And when finetuning on the downstream tasks, the model will output one of the representations $\h_{mol}$ or $\h_{fv}$ of a molecule to serve as learned molecular representation.
A downstream prediction head $f(\cdot)$ will use this representation as input, and predict the molecular property by $y=f(\h_{mol})$ or $y=f(\h_{fv})$.
% Here, the concepts should be clarified again: \textbf{fragment-pair view} refers to a fragment pair obtained after breaking an acyclic single bond of a molecule, and its representation and projection are $\h_{frag}^i$ and $\p_{frag}^i$, respectively.
% And \textbf{fragment view} refers to a new view that contains information of multiple views obtained after multi-view fusion mechanism, and its representation and projection are $\h_{fv}$ and $\p_{fv}$, respectively.

In addition, representations $\h_{frag}^i$ of $N_b$ fragment-pair views of a molecule goes through another projection head and a normalization layer to produce projection $\p_{frag}^i = norm(l_{frag}(\h_{frag}^i))$.
Inner product of these projections are computed to generate a similarity matrix $S=\{s_{ij} | s_{ij} = <\p_{frag}^i, \p_{frag}^j>\}, S \in \mathbb{R}^{N_b \times N_b}$, where $<\cdot,\cdot>$ denotes the inner product of two vectors.
The usage of this similarity matrix $S$ will be introduced in the next section.

% For pretraining, following the usage in MolCLR~\cite{}, the two projections $\p_{mol}$ and $\p_{fv}$ are used to calculate contrastive loss.
% And the similarity matrix $S$ will also be used to compute auxiliary similarity loss, which will be introduced in the next section.
% And when finetuning on the downstream tasks, the model will output the representations $\h_{mol}$ or $\h_{fv}$ of a molecule to serve as learned molecular representation.
% A downstream prediction head $f(\cdot)$ will use this representation as input, and predict the molecular property by $y=f(\h)$.

\subsection{Loss Functions}\label{sec:loss}
\begin{figure}[h]
\begin{center}
\centerline{\includegraphics[width = \linewidth]{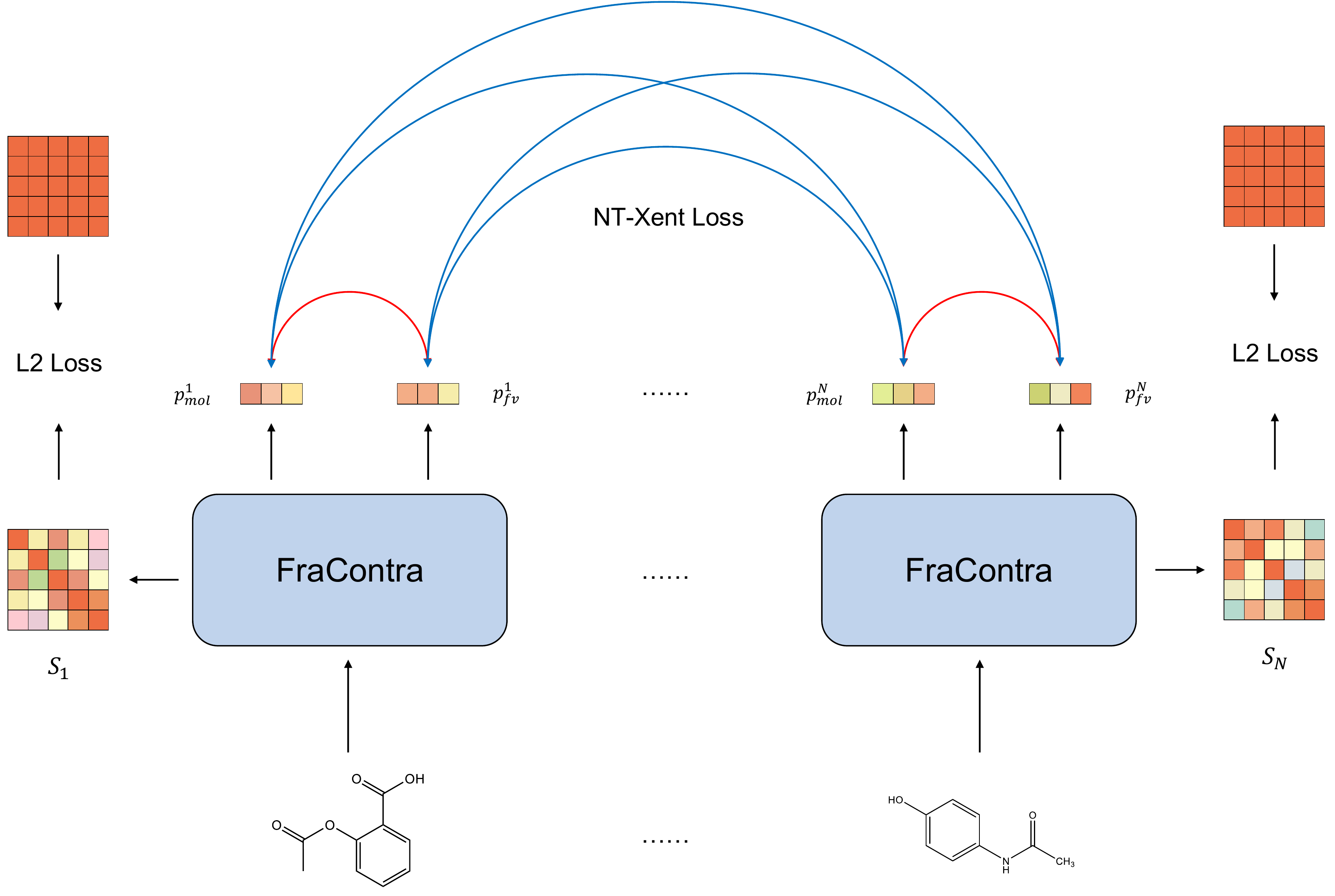}}
\caption{The illustration of FraSICL training. FraSICL is trained by both NT-Xent contrastive loss and an auxiliary similarity loss. In the contrastive loss, two projections of a molecule are treated as a positive pair, which is highlighted by red lines in the figure. Projections of other molecules in a batch are considered as negative pairs, which is shown by blue lines. For each molecule, L2 loss is computed between the similarity $S$ and an all-one matrix as auxiliary similarity loss.}
\label{fig:FraContra_Training}
\end{center}
\end{figure}

The training of FraSICL in the pre-training phase is illustrated in Fig.~\ref{fig:FraContra_Training}.
Here, given a batch of $N$ molecules, the model will calculate the projections $\p_{mol}$ and $\p_{fv}$ of each molecule.
Then, contrastive learning is performed between all samples in a batch.
The view pair (i.e. molecule view and fragment view) of each sample is a positive pair, as shown by the red line, and the view pairs of other samples in the batch are negative pairs, as shown by the blue line in the figure.
The NT-Xent Loss is used for contrastive learning:

\begin{equation}
% \resizebox{.91\linewidth}{!}{$
    \displaystyle
    \mathcal{L}_{mol}^{i} = \log \frac{\mathrm{e}^{\big(\frac{sim(\p_{mol}^i,\p_{fv}^i)}{\tau}\big)}}{\sum_{k=1}^{N}\mathbb{1}\{k \ne i\}\Big(\mathrm{e}^{\big(\frac{sim(\p_{mol}^i,\p_{mol}^k)}{\tau}\big)}+\mathrm{e}^{\big(\frac{sim(\p_{mol}^i,\p_{fv}^k)}{\tau}\big)}\Big)} \mbox{,}
    % $}
\end{equation}
\begin{equation}
% \resizebox{.91\linewidth}{!}{$
    \displaystyle
    \mathcal{L}_{fv}^{i} = \log \frac{\mathrm{e}^{\big(\frac{sim(\p_{mol}^i,\p_{fv}^i)}{\tau}\big)}}{\sum_{k=1}^{N}\mathbb{1}\{k \ne i\}\Big(\mathrm{e}^{\big(\frac{sim(\p_{fv}^i,\p_{mol}^k)}{\tau}\big)}+\mathrm{e}^{\big(\frac{sim(\p_{fv}^i,\p_{fv}^k)}{\tau}\big)}\Big)} \mbox{,}
        % $}
\end{equation}
where inner product similarity is adopted for $sim(\p_{mol}^i,\p_{fv}^i)$, and $\tau$ is a temperature parameter.
% Since each projection has been regularized and the length have been transformed to 1, at this time, the inner product similarity is equivalent to the cosine similarity.
The sum of all contrastive losses of a batch of molecules is denoted as $\mathcal{L}_{clr}=\sum_{i=1}^N \mathcal{L}_{mol}^i + \sum_{i=1}^N \mathcal{L}_{fv}^i$.

In addition, although the representations of different fragment-pair views have been fused, from the perspective of contrastive learning, the representations of different fragment-pair views of the same molecule should also be as close as possible.
And as demonstrated in previous study~\cite{zhang2021fragat}, representations of some fragment pairs of a molecule are highly predictive on the downstream tasks, while some others are less effective.
So, we hope that the representations of fragment-pair views can use information from each other to train the GNN encoder to extract better representations.
To this end, an additional auxiliary loss $\mathcal{L}_{sim}$ is introduced to improve the similarity between representations of fragment-pair views of a molecule, based on the similarity matrix $S$.
Since the inner product of two normalized vectors is equivalent to cosine similarity, and the maximum value of cosine similarity is 1, assuming a molecule $k$ have $N_b^k$ fragment-view pairs, the elements of the similarity matrix are $s_{ij}^k = <\p_{frag}^{k,i}, \p_{frag}^{k,j}>$, our auxiliary similarity loss of molecule $k$ is:

% In addition, since the model will produce different representations and projections for each fragment-pair view of a molecule.
% Base on an intuitive idea of contrastive learning, the representations of different views of the same sample should also be as close as possible.
% At the same time, the experiments in FraGAT~\cite{} have shown that representations of some fragment pairs of a molecule is highly predictive on downstream tasks, while representations of some other fragment pairs are less effective.
% So, we hope that the representations of fragment pairs can use information of each other to extract better representations.
% Therefore, although the FraContra model uses a fused fragment view which contains all of the information of the fragment pairs for contrastive learning, an additional auxiliary loss $\mathcal{L}_{sim}$ is introduced to improve the similarity between representations of fragment-pair views of a molecule.
% Since the inner product of two regularized vectors is equivalent to cosine similarity, and the maximum value of cosine similarity is 1, assuming a molecule $k$ have $N_b^k$ fragment pairs, the elements of the similarity matrix are $s_{ij}^k = <\p_{frag}^{k,i}, \p_{frag}^{k,j}>$.
% At this time, the calculation of the auxiliary similarity loss of sample $k$ is:

\begin{equation}\label{equ:similarity-loss}
    \mathcal{L}_{sim}^k = \frac{1}{(N_b^k)^2} \sum_{i=1}^{N_b^k} \sum_{j=1}^{N_b^k} (s_{ij}^k-1)^2 \mbox{,}
\end{equation}
i.e., as shown in Fig.~\ref{fig:FraContra_Training}, the sum of L2 loss between each element of the similarity matrix $S$ and that of an all-one matrix.
Denote the similarity loss of a batch of molecules as $\mathcal{L}_{sim}=\sum_{k=1}^N \mathcal{L}_{sim}^{k}$, then the loss for pre-training the FraSICL model is:

\begin{equation}
    \mathcal{L} = \gamma \mathcal{L}_{sim} + \mathcal{L}_{clr} \mbox{,}
\end{equation}
where $\gamma$ is a hyper-parameter to adjust the influence of the auxiliary similarity loss.

\section{Experiments}\label{sec:exp}

\subsection{Baseline experiments}
\textbf{Experimental setting.} To construct the pre-training dataset, 200K molecules are randomly sampled from the pre-training dataset of MolCLR, where 10 million molecules are gathered from the PubChem database~\cite{10.1093/nar/gkaa971}.
% 200K molecules are randomly sampled
% The pre-training dataset used in this paper is cited from MolCLR.
% The dataset used in MolCLR contains 10 million molecules, which are gathered from PubChem database.
% Here we randomly sample 200K molecules from the original 10 million dataset for training FraSICL.
The amount of pre-training data is generally smaller than that of the other baseline models, as shown in Tab.~\ref{tab:PTMpretraindetail}.
5\% of the pre-training data are randomly selected as a validation set for model selection.
% During training, 5\% of the samples in the pre-training dataset are used as a valid set to test the generalization ability of the pre-trained FraSICL model for model selection.
7 downstream tasks from MoleculeNet~\cite{wu2018moleculenet} are used as downstream target tasks for the baseline experiments.
% Pre-trained FraSICL model are finetuned on seven downstream tasks from MoleculeNet.
Scaffold splitting is used on each downstream task, with an $8:1:1$ ratio for the training/validation/test sets.

When transferring a pre-trained FraSICL model to the target tasks, different strategies can be applied, including using which branch of the model for producing molecular representations, and whether to finetune the pre-trained model (PTM) on target tasks.
% As mentioned before, the FraSICL model can output the representation of molecule view or fragment view for downstream MPP tasks, and one can also choose to freeze the PTM or finetune the PTM together with the downstream prediction head on target tasks.
In the baseline experiments, we adopt the more complex fragment view branch for molecular representations, and finetune the model together with prediction head on the target tasks.
% Due to the asymmetry of the FraSICL model, in the baseline experiments, we use the fragment view and finetuning PTM as experimental settings.
% And cases that using other settings will be discussed in the next section.

\textbf{Compared baseline models.} Seven state-of-the-art self-supervised pre-training models for molecular representation learning are used as baseline models for comparison, including MolCLR~\cite{wang2022molecular}, DMP~\cite{zhu2021dual}, MEMO~\cite{zhu2022featurizations}, GROVER~\cite{rong2020self}, GraphLoG~\cite{xu2021self}, PretrainGNNs~\cite{hu2019strategies} and KPGT~\cite{li2022kpgt}.
The experimental results are shown in Tab.~\ref{tab:exp}, where the data of baseline models are cited from the original papers of these models.
The best score on each dataset is bold, and the second-best is underlined.

\begin{table}[tbph]
\centering
\caption{The training details of the PTMs}
\begin{tabular}{lcr}
\toprule
PTM & Source of the pre-training dataset & Size \\
\midrule
MolCLR & PubChem & 10M \\
DMP & PubChem & 10M\\
MEMO & GEOM-Drug & 300K\\
GROVER & ZINC15 and ChEMBL & 11M\\
GraphLoG & ZINC15 & 2M\\
PretrainGNNs & ZINC15 (self-supervised) & 2M \\
& ChEMBL(supervised) & 456K\\
KPGT & ChEMBL & 2M\\
\midrule
FraSICL & PubChem & 200K\\
\bottomrule
\end{tabular}
\label{tab:PTMpretraindetail}
\end{table}

\begin{table*}[tbph]
    \centering
    \caption{Results of performance comparison between FraSICL and major existing models on 7 downstream MPP tasks.}
    \resizebox{\linewidth}{!}{
    \begin{tabular}{cccccccccc}
        \toprule
        Model & BACE & BBBP & ClinTox  & Tox21 & ESOL & FreeSolv & Lipop \\
         & classification & classification & classification & classification & regression & regression & regression\\
        \midrule
        MolCLR & 0.890 $\pm$ 0.003 & 0.736 $\pm$ 0.005 & 0.932 $\pm$ 0.017 &  0.798 $\pm$ 0.007 & 1.110 $\pm$ 0.010 & 2.200 $\pm$ 0.200 & 0.650 $\pm$ 0.080\\
        DMP-TF & 0.893 $\pm$ 0.009 & 0.781 $\pm$ 0.005 & \underline{0.950 $\pm$ 0.005} &  0.788 $\pm$ 0.005 & \underline{0.700 $\pm$ 0.084} & - & -\\
        MEMO & 0.826 $\pm$ 0.003 & 0.716 $\pm$ 0.010 & 0.816 $\pm$ 0.037 &  0.767 $\pm$ 0.004 & 0.984 $\pm$ 0.034 & - & 0.707 $\pm$ 0.001 \\
        GROVER & \underline{0.894 $\pm$ 0.028} & \underline{0.940 $\pm$ 0.019} & 0.944 $\pm$ 0.021 &  \underline{0.831 $\pm$ 0.025} & 0.831 $\pm$ 0.120 & \underline{1.544 $\pm$ 0.397} & \textbf{0.560 $\pm$ 0.035}\\
        PretrainGNNs & 0.845 $\pm$ 0.007 & 0.687 $\pm$ 0.013 & 0.726 $\pm$ 0.015 &  0.781 $\pm$ 0.006 & - & - & -\\
        GraphLoG & 0.835 $\pm$ 0.012 & 0.725 $\pm$ 0.008 & 0.767 $\pm$ 0.033 &  0.757 $\pm$ 0.005 & - & - & -\\
        KPGT & 0.855 $\pm$ 0.011 & 0.908 $\pm$ 0.010 & 0.946 $\pm$ 0.022 & \textbf{0.848 $\pm$ 0.013} & 0.803 $\pm$ 0.008 & 2.121 $\pm$ 0.837 & 0.600 $\pm$ 0.010\\
        \midrule
        FraSICL & \textbf{0.896 $\pm$ 0.010} & \textbf{0.948 $\pm$ 0.003} & \textbf{0.957 $\pm$ 0.011} &  0.807 $\pm$ 0.006 & \textbf{0.626 $\pm$ 0.008} & \textbf{1.094 $\pm$ 0.027} & \underline{0.581 $\pm$ 0.013}\\
        \bottomrule
    \end{tabular}
    }
    \label{tab:exp}
\end{table*}

\textbf{Results and analysis.} As shown in Tab.~\ref{tab:exp}, FraSICL achieves the best predictive performance on 5 of the 7 downstream MPP tasks, and the second on another one.
As the number of pre-training samples used by FraSICL is only 200K, which is the least among these compared baseline models, the experimental results show that FraSICL can make better use of the information contained in the graph fragments of molecules to produce molecular representations with better predictive performance.
Compared with the MEMO model that uses the same amount of pre-training data, the predictive performance of FraSICL on the 7 downstream tasks is significantly improved, even exceeds 20\% on the BBBP dataset.
And compared with the models such as GROVER and DMP-TF, FraSICL can achieve comparable or even higher predictive performance with only about $1/50$ training samples.
These results show the superiority of FraSICL to the existing models on molecular property prediction tasks.

\subsection{Experiments with different transferring settings}
\textbf{Experimental setting.} In the baseline experiments, we choose finetuning the more complex and predictive fragment view branch as the transferring setting.
% As mentioned in the previous section, in FraSICL, representations of molecule view or fragment view can be used for downstream molecular property prediction tasks, and one can choose to freeze the PTM or finetune the PTM on downstream tasks.
% Due to its asymmetric nature, in the previous section we choose to finetune the more complex and predictive fragment view branch of the model, and make prediction by the representation of fragment view.
In this section, other transferring settings are tested, i.e., the combinations of different branches and different fine-tuning strategies.
Experiments are carried out on the BBBP, ClinTox, ESOL and FreeSolv datasets.
Four transferring settings are evaluated, which are denoted as FraSICL-ft-mol, FraSICL-ft-frag, FraSICL-fr-mol, FraSICL-fr-frag, where \textit{ft} represents finetuning the PTM, \textit{fr} represents freezing the PTM, \textit{mol} indicates using molecule views and \textit{frag} indicates using fragment views.

\textbf{Results and analysis.} The experimental results are shown in Tab.~\ref{tab:transfer}.
Since the two branches of FraSICL are asymmetric, the structure of the fragment view branch is more complex and has stronger learning capability.
Thus, as is revealed by the experimental results, FraSICL-ft-frag achieves the best performance on 3 of the 4 target tasks.
% Thus, the FraSICL-ft-frag
% Among the four settings, the method which using fragment view and finetuning the PTM achieves the best performance on 3 of the 4 datasets.
% This result is as expected.
% Since the two branches of FraContra are asymmetric, the structure of the fragment view branch is more complex and have stronger learning capability, 
% thus it is expected to achieve the best performance.
% thus it achives the best performance on the 3 datasets.
% However, a more complex model structure indicates that it is more likely to face the risk of overfitting on downstream tasks
However, a more complex model structure indicates that it is more likely to suffer overfitting on the downstream tasks.
% Although the model has been trained on a large amount of unlabeled data and found a better initialization weight, the model will still face the risk of overfitting when there is too little data of downstream tasks.
So, when transferring to the FreeSolv dataset with only 642 samples, the performance of FraSICL-ft-frag is slightly inferior to that of FraSICL-ft-mol.
% So, it is revealed by the experimental results that, as the FreeSolv dataset has the least amount of data (only 642), the performance of FraSICL-ft-frag is slightly inferior to that of the FraSICL-ft-mol.
In addition, compared with freezing the pre-trained model, finetuning model allows the PTM to obtain information about specific molecular properties from the supervised loss, thereby the generated molecular representations are more relevant to the target task.
Thus a large improvement on the performance is achieved.

\begin{table}[tbph]
    \centering
        \caption{Predictive performance of 4 FraSICL model variants with different transferring settings on 4 MPP tasks.}
    % \resizebox{\linewidth}{!}{
    \begin{tabular}{cccccccccc}
        \toprule
        Model & BBBP & ClinTox &  ESOL & FreeSolv \\
         & classification & classification & regression & regression\\
        \midrule
        FraSICL-fr-mol & 0.852 $\pm$ 0.003 & 0.691 $\pm$ 0.007 & 1.321 $\pm$ 0.002 & 2.548 $\pm$ 0.006 \\
        FraSICL-fr-frag & 0.803 $\pm$ 0.010 & 0.628 $\pm$ 0.020 & 1.594 $\pm$ 0.217 & 2.293 $\pm$ 0.009\\
        FraSICL-ft-mol & 0.917 $\pm$ 0.006 & 0.906 $\pm$ 0.023 & 0.758 $\pm$ 0.029 & \textbf{1.085 $\pm$ 0.059}\\
        FraSICL-ft-frag & \textbf{0.948 $\pm$ 0.003} & \textbf{0.957 $\pm$ 0.011} & \textbf{0.626 $\pm$ 0.008} & 1.094 $\pm$ 0.027 \\
        \bottomrule
    \end{tabular}
    % }
    \label{tab:transfer}
\end{table}

\subsection{Influence of the auxiliary similarity loss}
\textbf{Motivation.} The auxiliary similarity loss, i.e., Equ.~(\ref{equ:similarity-loss}) introduced in Sec.~\ref{sec:loss}, is designed for making the fragment-pair views to learn from each other to better leverage the information encoded in different fragment-pair views.
% To let the representations of different fragment-pair views from the same molecule learn from each other, an auxiliary similarity loss $\mathcal{L}_{sim}$ is added to train a FraSICL model and the $\gamma$ is used to adjust the proportion of auxiliary loss in the final training loss function.
However, it is intuitive that when the auxiliary loss takes an excessively dominant role in the total training loss, the model may tend to generate
% However, it is intuitive that, though this loss is expected to help graph neural network to better encode graph fragment pairs, when it occupies an excessively large dominant role in the total training loss, the model may be trained to generate 
exactly the same representation vectors for different fragment-pair views to decrease the similarity loss.
On this occasion, the representation vectors of fragment-pair views will not contain any information about the topological structure, showing a model collapse phenomenon.
Thus, the influence of the hyperparameter $\gamma$ is crucial.
% And there is no need for the Transformer module to exist when the representations of different fragment-pair views are exactly the same.
% Therefore, to avoid model collapse and make both the graph encoder module and the Transformer module be well trained, the influence of the hyperparameter $\gamma$ is crucial.

\textbf{Experimental setting.} In this section, experiments are conducted to test the influence of the auxiliary similarity loss by setting different $\gamma$.
% to show the influence of the auxiliary similarity loss.
In these experiments, $\gamma$ is set to 0.1, 0.01, 0.005 and 0 respectively, where $\gamma=0$ indicates training without the similarity loss, which can be regarded as an ablation study.
% Except for $\gamma$, hyperparameters for pretraining the FraContra is set to be the same with those of the models that achieves the best predictive performance in Tab.~\ref{tab:exp}.
Here, the transferring setting is the same as the baseline experiments, i.e., finetuning the fragment view branch.
% During finetuning phase, the fragment view is used for prediction, and the pre-trained model is finetuned on downstream tasks.
% The same candidate hyperparameters for the prediction head are evaluated by grid search.

\begin{table}[tbph]
    \centering
        \caption{Experimental results on the influence of hyperparameter $\gamma$.}

    % \resizebox{\linewidth}{!}{
    \begin{tabular}{cccccccccc}
        \toprule
        $\gamma$ & BBBP & ClinTox &  ESOL & FreeSolv \\
         & classification & classification & regression & regression\\
        \midrule
        0.1 & 0.890 $\pm$ 0.023 & 0.916 $\pm$ 0.008 & 0.734 $\pm$ 0.028 & 1.151 $\pm$ 0.098\\
        % 0.05(15)\\
        0.01 & \textbf{0.948 $\pm$ 0.003} & \textbf{0.957 $\pm$ 0.011} & \textbf{0.626 $\pm$ 0.008} & \textbf{1.094 $\pm$ 0.027}\\
        0.005 & 0.927 $\pm$ 0.008 & 0.912 $\pm$ 0.012 & 0.667 $\pm$ 0.024 & 1.148 $\pm$ 0.015 \\
        0 & 0.906 $\pm$ 0.009 & 0.880 $\pm$ 0.030 & 0.662 $\pm$ 0.051 & 1.138 $\pm$ 0.092\\
        \bottomrule
    \end{tabular}
    % }
    \label{tab:gamma}
\end{table}

\textbf{Results and analysis.} Results are shown in Tab.~\ref{tab:gamma}. As shown in Tab.~\ref{tab:gamma}, the auxiliary similarity loss has an obvious impact on the predictive performance.
% In the four groups of experiments, as the transferring settings and the pretraining hyperparameters are exactly the same (except $\gamma$), it indicates that 
When $\gamma=0$, i.e., training without the similarity loss, the predictive performance is not superior to that of the other three models trained with the auxiliary similarity loss, which demonstrates that the auxiliary similarity loss can indeed promote the model to produce more predictive representations.
And when $\gamma=0.1$, the model achieves even worse results than $\gamma=0$ on 3 of the 4 downstream tasks, which reveals that a large value of $\gamma$ may make the similarity loss be harmful to the model and lead to performance degradation.
% that when the value of $\gamma$ is too large, the effect of the auxiliary similarity loss may be harmful for the model which may lead to a performance degradation.

\section{Conclusion}
This paper focuses on the semantic inconsistency problem that may occur when using noise-adding operations to generate new views for contrastive learning in self-supervised molecular property prediction studies.
To solve this problem, this paper first defines semantic-invariant molecular view by introducing two types of semantic inconsistent views that may lead to false positive pairs and consequently poor performance.
% first formally defines two types of semantic inconsistency problem in the context of molecular representation contrastive learning and points out that the semantic inconsistency problem may lead to false positive pairs.
Then, a semantic-invariant view generation method is proposed.
The views generated by this method will not cause semantic inconsistency, which realizes the decoupling of the input graph topology and the message passing topology of GNNs.
Thus, this method is expected to promote the GNN encoders to extract better molecular representations.

Based on the semantic-invariant views, a Fragment-based Semantic-Invariant Contrastive Learning (FraSICL) molecular representation model is developed.
FraSICL is an asymmetric model with two branches, the molecule view branch and the fragment view branch.
A multi-view fusion mechanism is also introduced to make better use of the information contained in the views of different fragment pairs. Furthermore, an auxiliary similarity loss is designed to train the backbone GNN to produce better representations.
% In the fragment view branch, a graph neural network is used to extract the representations of fragment-pair views of a given molecule, and a Transformer is used for multi-view fusion to make full use of the information of each fragment-pair view and focus on the more important fragment-pair view to generate better representations of fragment view.
% The representations of molecule view and fragment view are used as positive pairs for contrastive learning, and an auxiliary similarity loss is added to make
% better use of the information contained in different views to produce informative representations.

Baseline experiments are conducted on 7 target tasks, and experimental results show that FraSICL achieves state-of-the-art predictive performance with the least number of pre-training data.
Further experiments demonstrate that in our model finetuning is effective in boosting performance and the auxiliary similarity loss can improve the predictive accuracy if a proper hyperparameter $\gamma$ is selected.
% a subtle portion of similarity loss will improve the 
These findings reveal that FraSICL can make better use of the information of pre-training samples and generate representations with superior predictive performance.

%Bibliography
\bibliographystyle{unsrt}  
\bibliography{references}

\end{document}